# Stabilization Analysis and Mode Recognition of Kerosene Supersonic Combustion: A Deep Learning Approach Based on Res-CNN-β-VAE

Weiming Xu[a], Tao Yang[a], Chang Liu[b], Kun Wu[b*], Peng Zhang[a*]

[a] *Department of Mechanical Engineering, City University of Hong Kong, Kowloon Tong, Kowloon, 999077, Hong Kong*
[b] *State Key Laboratory of High Temperature Gas Dynamics, Institute of Mechanics, Chinese Academy of Sciences, Beijing, 100190, People's Republic of China*

[*]*Corresponding Author Email: Kun Wu, wukun@imech.ac.cn; Peng Zhang, penzhang@cityu.edu.hk;*

**Abstract:** The scramjet engine is a key propulsion system for hypersonic vehicles, leveraging supersonic airflow to achieve high specific impulse, making it a promising technology for aerospace applications. Understanding and controlling the complex interactions between fuel injection, turbulent combustion, and aerodynamic effects of compressible flows are crucial for ensuring stable combustion in scramjet engines. However, identifying stable modes in scramjet combustors is often challenging due to limited experimental measurement means and extremely complex spatiotemporal evolution of supersonic turbulent combustion. This work introduces an innovative deep learning framework that combines dimensionality reduction via the Residual Convolutional Neural Network-$\beta$-Variational Autoencoder (Res-CNN-$\beta$-VAE) model with unsupervised clustering (K-means) to identify and analyze dynamical combustion modes in a supersonic combustor. By mapping high-dimensional data of combustion snapshots to a reduced three-dimensional latent space, the Res-CNN-$\beta$-VAE model captures the essential temporal and spatial features of flame behaviors and enables the observation of transitions between combustion states. By analyzing the standard deviation of latent variable trajectories, we introduce a novel method for objectively distinguishing between dynamic transitions, which provides a scalable and expert-independent alternative to traditional classification methods. Besides, the unsupervised K-means clustering approach effectively identifies the complex interplay between the cavity and the jet-wake stabilization mechanisms, offering new insights into the system's behavior across different gas-to-liquid mass flow ratios (GLRs).

## Introduction

The supersonic ramjet engine (scramjet) is a key propulsion system for hypersonic vehicles and utilizes supersonic airflow to achieve both gas compression and combustion, eliminating the necessity for onboard oxidizers [1]. Its simple design and potential for high efficiency make it a promising technology for aerospace applications. Enhancing scramjet performance over a wide operational envelope, especially in terms of combustion stability, is essential to satisfying the maneuverability requirements of future hypersonic vehicles.

The stability of supersonic turbulent combustion is significantly affected by the specific behavior or characteristics of flames, referred to as the flame mode. In scramjet combustors, the incoming airflow velocity typically surpasses the flame propagation speed, resulting in an extremely short residence time for the fuel-air mixture. This leads to the occurrence of corresponding flame stabilization mechanisms, such as cavity flame stabilization and jet wake flame stabilization modes, and transitional mode between these states [2], to achieve a sustainable combustion. These dynamical combustion processes in distinct modes are influenced by various factors such as combustion chamber design, flow dynamics, and operating conditions. Identifying and understanding these modes is essential for optimizing scramjet design and ensuring stable performance under varying operational conditions.

Furthermore, distinguishing the transition and stable processes of combustion and characterizing various stable flame modes in scramjet combustors presents formidable challenges, primarily due to the limited experimental measurement means and highly complex spatial-temporal dynamics involving multi-scale and non-linear interactions between flow and chemistry [3]. Various conventional approaches for mode recognition are limited to their empirical and non-general characteristics. Therefore, advanced spatial-temporal analysis techniques capable of efficiently capturing the essential features of scramjet combustion are becoming increasingly vital. This has led to growing interest in reduced-order models (ROMs) that maintain key information in a small dimensional space. The reduced-order techniques are typically classified into linear and nonlinear approaches [4]. Proper Orthogonal Decomposition (POD) is one of the most widely used techniques in combustion, fluid dynamics, and industrial applications, valued for its capability to discern dominant modes of variation in a given dataset and projecting the data onto a lower dimensional subspace spanned by these key modes, making it an achievable tool for analyzing complex, high-dimensional systems such as scramjet combustors [5]. Another well-established linear method is Dynamic Mode Decomposition (DMD), which identifies dynamical modes driving the temporal evolution of the supersonic combustion systems [6]. Although POD, DMD, and their variants have proven effective in reducing the dimensionality of some combustion problems, their reliance on linear bases poses limitations in capturing the complex nonlinear interactions inherent in the high-speed supersonic flow and combustion.

Recently, nonlinear ROMs, particularly machine learning-based methods, have gained attention for their ability to model complex dynamics [7]. Autoencoder neural networks, such as Autoencoders (AEs), Variational Autoencoders (VAEs), and Beta-VAE ($\beta$-VAEs), offer an effective way to capture nonlinear representations in the latent space through their encoder-decoder architecture [8]. The latent representations can be utilized for mode recognition via classification methods, such as K-means clustering [9], and model establishment for describing the dynamic state of the complex system, such as the SINDY approach [10].

The $\beta$-VAEs modify the autoencoder neural network with an adjustable hyperparameter $\beta$ for learning a compact latent representation, as it can encode distributions rather than points within a structured latent space, with the $\beta$ parameter balancing reconstruction accuracy, regularization, and latent-space disentanglement. The powerful framework has been proven to be effective in investigating complex, high-dimensional rocket combustion [11]. The regularization for ensuring smooth and structural representations, continuity for enabling interpolation between modes, and a probabilistic structure for handling uncertainty and complex dependencies in the latent space [12] makes the $\beta$-VAEs well-suited for analyzing complex dynamic combustion modes.

Images are a prevalent type of data in experiments and simulations. Convolutional Neural Networks (CNNs) are particularly adept at capturing intricate spatial patterns within image data [13]. Nevertheless, CNN-based models often demonstrate decreased accuracy of reconstruction due to the vanishing gradient problem, especially when addressing complex phenomena such as turbulent flows [14]. To mitigate this problem, architectures like Residual Networks (ResNet) [15] can allow for deeper networks to effectively extract multi-scale spatial features essential for dynamic mode identification and analysis in complex flow systems. In recent years, Guo et al. [16] utilized an advanced ResNet-based CNN to perform high-fidelity, super-resolution reconstruction of combustion flow fields. Zhou et al. [17] used a ResNet-CNN to process flame images and reached a high prediction accuracy of thermoacoustic state.

As the snapshot analysis of CH* chemiluminescence from the supersonic model combustors was carried out for identifying various dynamical states, we introduce the ResNet-CNN into the $\beta$-VAEs approach, short for Res-CNN-$\beta$-VAE. This approach combines ResNet-based CNN blocks for capturing intricate spatial features with a $\beta$-VAE for modelling meaningful latent representations, enabling a comprehensive and insightful analysis of the complex combustion system. Next, unsupervised mode recognition is achieved based on the phase distribution of flame dynamical processes in the latent space through a K-means clustering algorithm. Furthermore, the recognition performance is evaluated with multiple metrics targeting specific and overall aspects, such as accuracy (AC), precision (PR), and area under the curve (AUC) [8].

**Methods**

As shown in Fig.1, the supersonic combustion experiment was conducted in the same combustion test facility as described in our previous work [18]. An overview description is given here. To simulate Mach number 6 flight conditions corresponding to a Mach 2.5 inlet flow at the combustor isolator, the experiment maintained a total temperature of $1600 \pm 20$ K and total pressure of $0.93 \pm 0.02$ MPa. The mass flow rate of vitiated air was maintained at $1.0 \pm 0.02$ kg/s. Liquid RP-3 kerosene with air-assisted atomization was injected through an orifice of 2.1 mm in diameter at the streamwise location 56 mm upstream of the cavity leading edge. The flame stabilization modes could be modulated by varying gas-to-liquid mass flow ratios (GLRs) while the overall equivalence ratio was fixed at 0.6. During the experiments, the pilot hydrogen was initially injected into the combustor to assist ignition and turned off at 6.5 s, while kerosene was introduced at 5.5 s and burned intensely in the time range of 7.0 to 7.5 s according to the measurement of pressures for typical streams.

To facilitate stability analysis of the reaction zone during combustion, CH* chemiluminescence was measured to monitor the flame dynamics by adopting a high-speed camera (10,000 fps, 30 μs exposure, 20 μm/pixel) and a CH* optical filter. To handle time-series snapshots of CH* chemiluminescence (the high-dimensional, spatial-temporal data) from the supersonic combustion experiments, the Res-CNN-$\beta$-VAE (Residual Convolutional Neural Network-$\beta$-Variational Autoencoder) was designed and established by integrating three distinct ResNet-based CNN blocks (ResCNN) for spatial feature extraction, as shown in Fig. 2. Specifically, each ResCNN block includes batch normalization, convolutional layers, and Exponential Linear Unit (ELU) activation functions, with the ResNet UpCNN blocks additionally featuring upsampling layers. These spatial feature modules are coupled with a $\beta$-VAE for probabilistic latent modelling. Generally, there are seven components in the present architecture. Component 1 (Input): the input data are the sequential CH* chemiluminescence snapshots with 560 × 240 pixels and 9003 time samples; Component 2 (CNN blocks): the CNN blocks are designed as a CNN layer (1-64 channels) and a ResNet CNN block (64-64 channels) for feature extraction, four ResNet DownCNN blocks (64-128-256-512-1024 channels) for progressive spatial resolution reduction with a Sigmoid activation layer for nonlinear mapping, and followed by a CNN layer (1024-256 channels) to compress the feature maps;

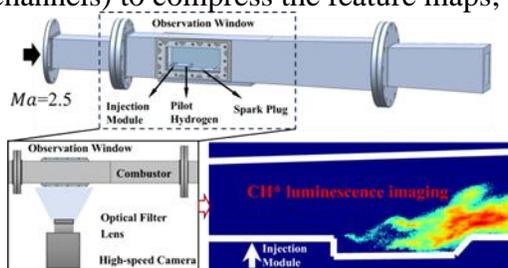

Fig.1. Schematic diagram of the combustion chamber with a flame-holder cavity module and devices for the CH* chemiluminescence in the observation window.

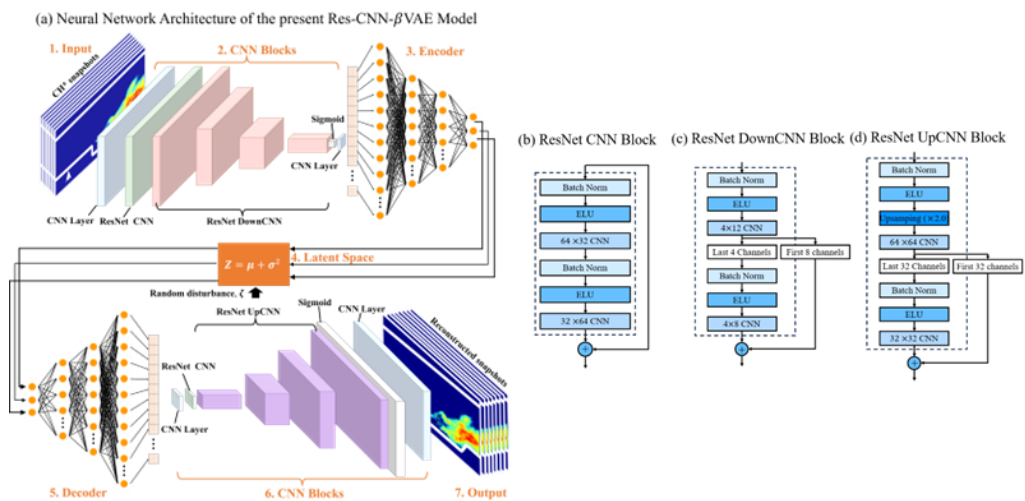

Fig. 2. (a) Neural network architecture of the present Res-CNN-$\beta$-VAE model, featuring a variational autoencoder with encoder, latent space, and decoder, along with three CNN blocks: (b) ResNet CNN, (c) ResNet DownCNN, and (d) ResNet UpCNN.

The $\beta$-VAE contains Component 3 (Encoder), Component 4 (Latent space), and Component 5 (Decoder), all designed with fully connected layers for dimensionality reduction, learned latent representations, and data generation, respectively; Component 6 (CNN blocks): the CNN blocks following the decoder of the $\beta$-VAE consist of a CNN layer (256-1024 channels) and a ResNet CNN block (1024-1024 channels) for hierarchical feature extraction, followed by four ResNet UpCNN blocks (1024-512-256-128-64 channels) with upsampling layers to progressively increase spatial resolution, with a CNN Layer (64-1 channel) to facilitate latent space reconstruction. A Sigmoid activation layer introduces nonlinearity, while the final CNN layer refines the output for data generation; Component 7 (Output): the outputs are the reconstructed CH* chemiluminescence snapshots.

The Res-CNN-$\beta$-VAE model and training pipeline were implemented using the Torch 1.12 deep learning framework. Training was conducted on NVIDIA A800 GPUs, taking approximately 20 hours. The hyperparametric value $\beta$ was optimized to $10^{-4}$ to balance the reconstruction accuracy and latent-space disentanglement through extensive experimentation. The model was optimized using the Adam algorithm with a learning rate of $1 \times 10^{-5}$. The training strategy uses an early stopping to prevent overfitting, terminating after 100 epochs of no improvement in validation loss or once the predefined epoch limit is reached. The model underwent 1000 epochs with a batch size of 100 and contained $5.35 \times 10^7$ trainable parameters. Training used the first 80% of the time-series snapshots, with the remaining 10% reserved for validating and 10% reserved for testing

**Results and Discussion**
**Dynamical Behavior of Supersonic Combustion**

Figure 3 depicts the fully developed state for combustion stabilization after withdrawing the pilot hydrogen and the probability density distribution of the flame front positions in a scramjet combustor under different GLRs. The CH* chemiluminescence images within about 1.0 s in three cases of GLR=10%, 14%, and 20% show that flames progressively achieve a stable state. However, their flame dynamical behaviors are distinct with varying GLRs. Specifically, as the GLR increases, the influence of the fuel flow on combustion dynamics becomes more pronounced, generally resulting in three distinct dynamical modes.

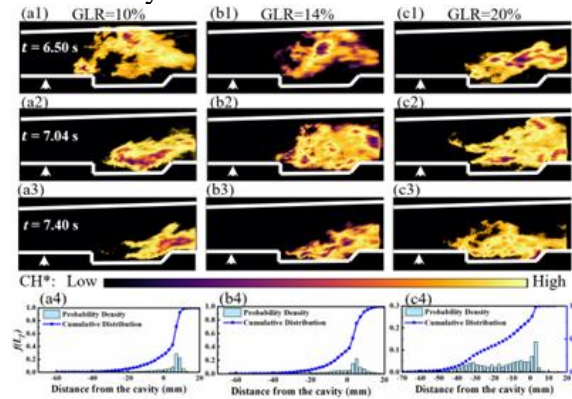

Fig. 3. Experimental results of supersonic combustion at statistically stable state after withdrawing the pilot hydrogen and the probability density distribution of the flame front positions under GLR= (a1-a4) 10%, (b1-b4) 14%, and (c1-c4) 20%. The contours indicate the CH* distribution and the histograms show the positional distribution of flame fronts.

Mode I: at GLR=10%, the combustion system is primarily stabilized within the cavity and its shear layer, as shown in Fig. 3(a1-a3). The flame front remains largely confined to the recirculation zones, with most combustion (flame fronts) occurring within or just downstream of the cavity in Fig. 3(a4).

Mode II: at GLR=14%, the enhanced atomization and penetration of the fuel jet cause the flame front to move upstream, extending beyond the cavity in Fig. 3(b1-b3). This shift marks a growing influence of the fuel jet on flame stabilization, resulting in a broader and more dynamic reaction zone, while the cavity associates flame stabilization in Fig. 3(b4).

Mode III: at GLR=20%, a significant portion of the flame extends into the jet region, with some flame still anchored behind the cavity, as shown in Fig. 3(c1-c3). However, the combustion system becomes increasingly dominated by the jet wake zones, resulting in a more diffuse and unstable flame front in Fig. 3(c4). While the cavity continues to contribute to stabilization, the system exhibits greater fluctuations in flame position and intensity, featuring the jet-dominated combustion behavior in Mode III. This progression underscores the growing influence of the jet wake, leading to more complex and unpredictable combustion behaviors, marked by increased instability and dynamic fluctuations.

**Identification of Flame Transition to Stable State**

Figure 4 illustrates phase point distributions of supersonic combustion dynamics in the three GLR cases in the phase space of three latent variables of $Z_1$, $Z_2$, and $Z_3$, outputting from the encoder of the Res-CNN-$\beta$-VAE. The motivation for using a three-dimensional latent space was derived from the physical consideration that the dynamical processes in supersonic combustion with twin-fluid injection intrinsically exhibit strong turbulent features, where three-dimensional interactions of vortices occur essentially. Besides, the present results show that the three latent variables are sufficient for mode recognition, although we are aware that a higher-dimensional latent space may be beneficial for obtaining a finer representation of the dynamical systems.

This latent space captures critical spatial-temporal features of the high-dimensional data of supersonic combustion, effectively mapping the evolution of flame behaviors under different GLRs. In all three cases, phase trajectories exhibit consistency in the evolution direction, which somehow manifests the dynamical transition pathways from an unstable state (blue lines in Fig. 4) to a stable state (red lines in Fig. 4) within the period of 6.5~7.4 s. Furthermore, the phase trajectory in each case has two bunches of dense scatters (blue and red phase points) in the two-dimensional latent spaces of each two of $Z_1$, $Z_2$, and $Z_3$, implicitly revealing that distinct clusters of a phase trajectory are associated with different combustion states, highlighting the ability of the latent space to distinguish between unstable and stable combustion behaviors. Besides, we noted that these transitions are particularly sensitive to GLR, as higher GLRs exhibit smoother trajectories with reduced variability, while lower GLRs lead to more erratic transitions prior to the dynamically stable combustion.

At lower GLRs in Fig. 4(a) and 4(b), the phase point distributions show significant variability, indicating that the combustion system remains sensitive to transient fluctuations caused by shutting down the pilot hydrogen. In this regime, flame stabilization is primarily controlled by the cavity and its shear layer. The scattered phase trajectories indicate an unstable combustion state, driven by transient fluctuations following pilot hydrogen shutdown. As the system transitions toward a more stable state, driven by the stabilization mechanism within the cavity, the trajectories gradually narrow. At a higher GLR in Fig. 4(c), the phase trajectory has reduced variability, compared to lower GLR cases, reflecting a diminished sensitivity to hydrogen shutdown. This is due to the increasing dominance of jet wake-driven combustion, which induces greater instability and dynamic fluctuations. Enhanced atomization gas flow improves atomization and penetration, shifting flame stabilization toward the jet wake. This results in more pronounced flame front oscillations, driven by interactions between the fuel jet and wake recirculation zones, reflected as more erratic and fluctuating trajectories in latent space, corresponding to greater fluctuations in combustion behavior. This shift underscores the growing influence of the jet wake over the cavity as GLR increases, leading to more complex and unstable combustion dynamics.

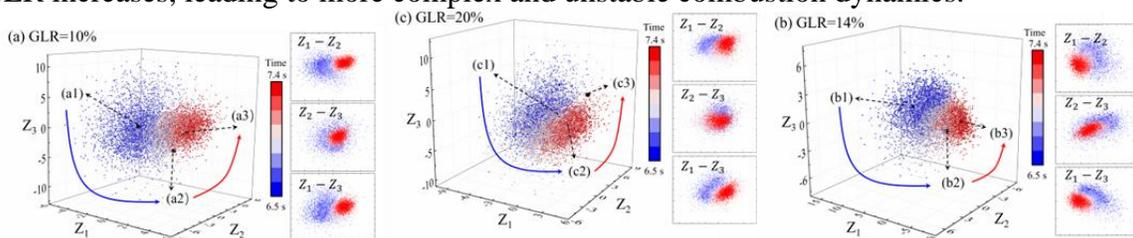

Fig. 4. Phase distributions in three-dimensional latent space of the present Res-CNN-$\beta$-VAE model for supersonic combustion process from unstable state with pilot hydrogen to stable state after pilot hydrogen shutdown under GLR= (a) 10%, (b) 14%, and (c) 20%.

By analyzing the time evolution of latent variables, we provide further illustrations of the flame dynamics observed in the phase space. For the cases of 10% and 14% GLRs in Fig. 5(a) and 5(b), it can be observed that $Z_1$, $Z_2$, and $Z_3$ oscillate significantly with time and all exhibit similar trends with the variance change of gradually being narrow. To further validate this finding, the standard deviation of the variable $\Theta = \sqrt[3]{Z_1 Z_2 Z_3}$ is plotted to illustrate the global trend, as shown in Fig. 5(d). Based on the domain knowledge of flame physical evolution from the unstable state to the stable state, we infer that the narrow trend manifests the convergence of flame dynamics and the regime of the small variance of the latent variables corresponding to the flame stable state. Furthermore, we observed that the higher GLR is, the later the flame stabilization occurs, reflecting the increased complexity in the combustion process as higher GLR leads to more dynamic interactions between the fuel jet and the recirculation zones. In particular, the convergence trend of variance of latent variables is less obvious in the 20% GLR case in Fig. 5(c), with increased erratic fluctuations during the dynamic process. The variance is harder to converge at a relatively larger GLR, aligning with the stronger turbulent and oscillatory combustion.

Hereto, we have carried out the analyses of phase distribution and time-varying evolution of three latent variables, offering a comprehensive and robust framework for understanding the dynamic transitions between instability and stability in scramjet combustion. This approach not only provides deeper insights into combustion dynamics but also eliminates the subjectivity inherent in traditional expert-driven methods, paving the way for more precise and scalable analyses of complex combustion systems.

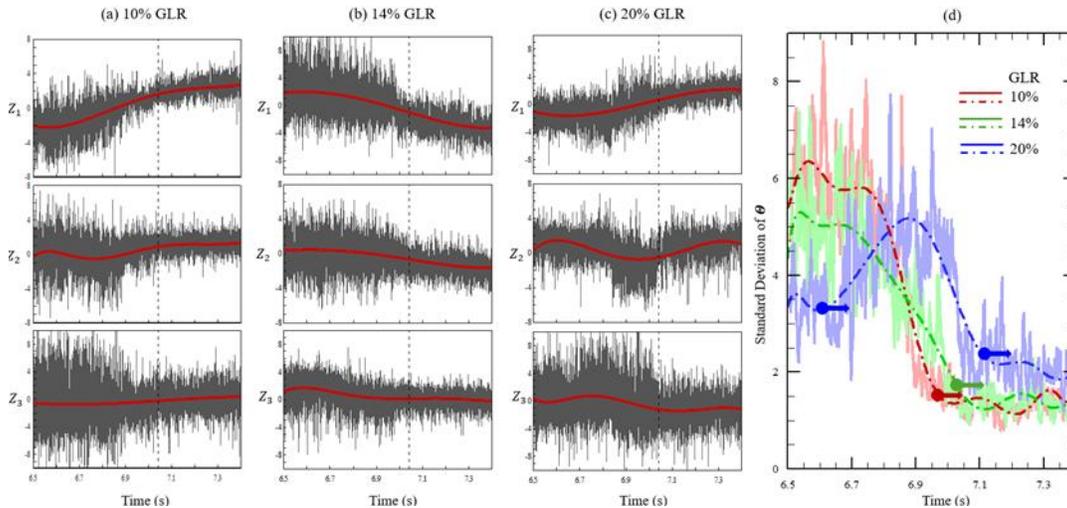

Fig. 5. Time-varying evolution of latent variables $Z_1$, $Z_2$, and $Z_3$, in the cases of (a) 10%, (b) 14%, and (c) 20% GLRs. The red lines are the trendline of time-average value. The dotted lines are the stabilization onset time observed by the conventional approach. (d) The standard deviation of the variable $\Theta = \sqrt[3]{Z_1 Z_2 Z_3}$ vs. time for identifying the transition of dynamical states in the three cases. The arrows indicate the critical points for remarkable changes in flame dynamical states.

### 3.3 Unsupervised Mode Recognition of Supersonic Combustion

To distinguish flame-stable states for different GLRs, we investigate the three modes for mode recognition by an unsupervised approach. Utilizing the K-means algorithm [9], a robust unsupervised classification approach without labeling data in advance, we classify these modes based on phase distributions of three latent variables via the Res-CNN-$\beta$-VAE. The deep learning framework of integrating dimensionality reduction and unsupervised classification is, for the first time, proposed for mode recognition of flame modes in supersonic combustors. It could be easily transferred to explore massive raw data (e.g., experimental or computational snapshots).

As illustrated in Section 3.2, we have recognized Mode I, for example, the 10% GLR case, where combustion is primarily stabilized within the cavity; Mode II, for example, the 14% GLR case, in which the jet wake starts to affect the flame dynamics while the flame stabilization is still assisted by the cavity; Mode III, for example, the 20% GLR case, where the cavity continues to contribute to stabilization but jet-dominated stabilization works as well.

Taking an example of 1000 snapshots in each flame stable state, we classify the cases of 10% and 20% GLRs in an unsupervised way and have mode recognition results, as shown in Fig. 6. As a comparison, the ground truth of phase distribution serves as a baseline in Fig. 6(a). The predicted phase points classified into distinct flame modes via the K-means algorithm are shown in Fig. 6(b). The confusion matrix (a prediction summary) in Fig. 6(c) shows that the present approach works well in general.

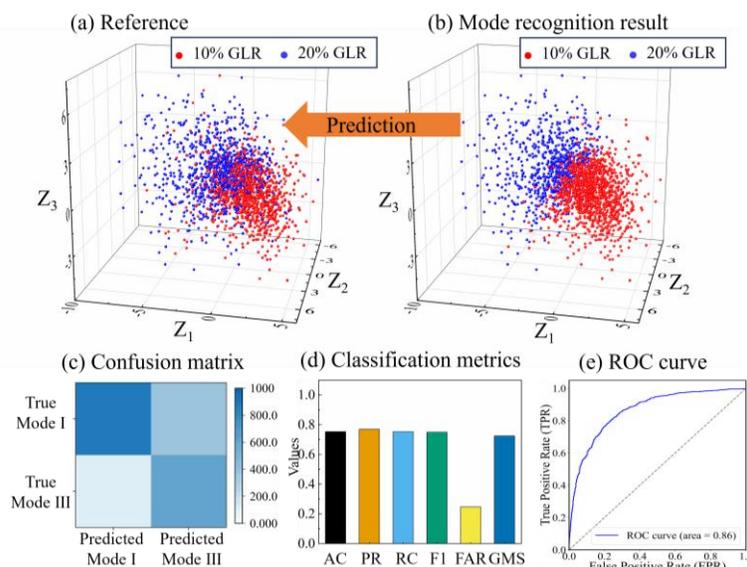

Fig. 6. Unsupervised classification of stable combustion modes and performance evaluation: (a) ground truth phase distribution, (b) prediction results of K-means classification, (c) confusion matrix, (d) classification performance metrics, (e) receiver operating characteristic (ROC) curve.

**Conclusions**

In this study, we propose a novel deep learning approach for analyzing supersonic combustion dynamics through latent space mapping, combining the CNN-$\beta$VAE model for dimensionality reduction with K-means clustering for unsupervised classification. The data are two-dimensional CH* chemiluminescence snapshots collected from supersonic combustion experiments.

By reducing high-dimensional combustion data to a three-dimensional latent space, we capture the essential spatial-temporal dynamics of flame behavior under varying gas-to-liquid mass flow ratios (GLRs). This approach enables the observation of dynamic combustion transitions, effectively mapping the combustion dynamical process in physical space into low-dimensional latent space.

By analyzing the variance trends of time-varying latent variables, we propose a Res-CNN-$\beta$-VAE method to distinguish dynamic transitions, offering a scalable and objective alternative to traditional expert-driven classifications. The K-means clustering further identifies distinct flame modes. The unsupervised results show that the proposed approach has nice performance and broader application potential in distinguishing flame modes.

In addition, we have fully realized that the Res-CNN-$\beta$-VAE model, treating each snapshot independently and only considering the temporal sequence as ordered inputs, struggles to capture the deep temporal dependencies inherent in combustion. In future work, we will focus on integrating advanced time-series models, such as Transformers [20], to enhance the capture of temporal dependencies and improve the depiction of time-varying combustion dynamics.

**Acknowledgements**

The authors declare that they have no known competing financial interests or personal relationships that could have appeared to influence the work reported in this paper.